\documentclass[aps,prl,reprint,groupedaddress,bibnotes,superscriptaddress,showpacs]{revtex4-1}
\usepackage{graphicx}
\usepackage{color}
\DeclareTextFontCommand{\emph}{\bfseries}
\begin{document}
%\preprint{}
\title{Long lifetimes in optical ion traps}
\author{Alexander Lambrecht}\thanks{These authors contributed equally. Corresponding author: julian.schmidt@physik.uni-freiburg.de}\affiliation{Albert-Ludwigs-Universit\"{a}t Freiburg, Physikalisches Institut, Hermann-Herder-Stra{\ss}e 3, 79104 Freiburg, Germany}
\author{Julian Schmidt}\thanks{These authors contributed equally. Corresponding author: julian.schmidt@physik.uni-freiburg.de}\affiliation{Albert-Ludwigs-Universit\"{a}t Freiburg, Physikalisches Institut, Hermann-Herder-Stra{\ss}e 3, 79104 Freiburg, Germany}
\author{Pascal Weckesser}
\author{Markus Debatin}
\affiliation{Albert-Ludwigs-Universit\"{a}t Freiburg, Physikalisches Institut, Hermann-Herder-Stra{\ss}e 3, 79104 Freiburg, Germany}
\author{Leon Karpa}\thanks{These authors contributed equally. Corresponding author: julian.schmidt@physik.uni-freiburg.de}\affiliation{Albert-Ludwigs-Universit\"{a}t Freiburg, Physikalisches Institut, Hermann-Herder-Stra{\ss}e 3, 79104 Freiburg, Germany}
\affiliation{Freiburg Institute for Advanced Studies (FRIAS), Albert-Ludwigs-Universit\"{a}t Freiburg, 79104 Freiburg, Germany}
\author{Tobias Schaetz}
\affiliation{Albert-Ludwigs-Universit\"{a}t Freiburg, Physikalisches Institut, Hermann-Herder-Stra{\ss}e 3, 79104 Freiburg, Germany}


\date{\today}
\begin{abstract}

We report on single Barium ions confined in a near-infrared optical dipole trap for up to three seconds in absence of any radio-frequency fields.
Additionally, the lifetime in a visible optical dipole trap is increased by two orders of magnitude as compared to the state-of-the-art using an efficient repumping method.
We characterize the state-dependent potentials and measure an upper bound for the heating rate in the near-infrared trap.
These findings are beneficial for entering the regime of ultracold interaction in atom-ion ensembles exploiting bichromatic optical dipole traps. 
Long lifetimes and low scattering rates are essential to reach long coherence times for quantum simulations in optical lattices employing many ions, or ions and atoms.
\end{abstract}

%\keywords{}
\maketitle

% the file 'lifetime_wordcount_bodyonly' can be used for word counting with the command ./wordcount.sh. It does not include abstract, authors, title, acknoledgements and bibliography. There are currently (15.09.2016 at 18h31) ~ 3311 words + figures. Figures count as 150/aspectratio + 20. Aspect ratios of the figures are:
% figure1 = 561/272 = 2.06 ; words = 73
% figure2 = 202/128 = 1.58 ; words = 95
% figure3 = 200/127 = 1.57 ; words = 95
% figure4 = 200/127 = 1.57 ; words = 95
% total = 358 words
% combined : 3670 words, PRL limit is 3750, awesome! (15.09.2016 at 18h31)
%
Radio-frequency (rf) traps for atomic ions feature deep trapping potentials and long lifetimes, as well as long range Coulomb interaction.
For decades, advanced tools for the manipulation of trapped ions have been developed and are propelling numerous fields of research \cite{Paul1990}.
For example, progress in the realm of quantum information processing permits unique individual control of motional and electronic states in strings of up to $10$ ions.
Operational fidelities close to unity for state preparation, phonon-mediated interactions and detection allow for reaching or setting the state-of-the-art in quantum metrology, quantum computation and experimental quantum simulation \cite{Chou2010,Wineland2013,Debnath2016,Monz2016}.
However, trapping ions in rf fields implicates inevitable effects, such as rf-driven motion.
This micromotion, which is superimposed on the secular motion of the ion in its time-averaged trapping potential, is undesirable or detrimental to many applications \cite{Chou2010,Cetina2012,Schneider2012b,Thompson2015}.
In addition, scaling Coulomb crystals in size and dimension while preserving the level of control over motional and electronic states remains a demanding task \cite{Kielpinski2002,Home2009,Mielenz2016}.

To address questions of many-body physics in the quantum regime, it has been proposed to spatially overlap ions in rf traps and neutral atoms in optical fields \cite{Zoller2000,Doerk2010}. This has led to seminal experiments in so-called hybrid traps \cite{Grier2009,Zipkes2010,Kruekow2016,Meir2016}.
However, it has been observed that the temperature of an ion embedded in a cloud of ultracold atoms is limited to temperatures on the order of $1\,\text{mK}$ independent of the temperature of the bath of typically $100\,\text{nK}$ \cite{Kruekow2016}.
These findings are in agreement with theory, which predicts the collision energy to be fundamentally dominated by rf micromotion induced during the atom-ion interaction, even under the assumption of zero stray electric field \cite{Cetina2012,Krych2015}.

Recently, atomic ions have also been confined by optical fields using different approaches.
Superimposing a one-dimensional optical lattice on radial rf and axial electrostatic (dc) confinement of a linear Paul trap \cite{Linnet2012,Karpa2013} has paved the way for novel applications, such as the study of classical and quantum phase transitions in the context of friction \cite{Bylinskii2015}.
Trapping $^{24}\text{Mg}^{+}$ ions in optical fields without rf confinement has been achieved both in a near-resonant optical dipole trap \cite{Schneider2010} and in a one-dimensional optical lattice \cite{Enderlein2012}.
In principle, optical potentials allow for versatile trapping geometries scalable to two and three dimensions and are routinely used for neutral particles \cite{Bloch2008}.
However, the lifetime of the ions in the dipole trap was limited to milliseconds due to photon recoil heating caused by off-resonant scattering from the dipole laser.
Recently, trapping of $^{138}\text{Ba}^{+}$ ions in a far-detuned optical dipole trap at $532\,\text{nm}$ has been reported \cite{Huber2014}.
Despite a reduction of the scattering rate by three orders of magnitude, the lifetime remained limited to a few milliseconds. This was assumed to be due to residual scattering into metastable electronic states featuring repulsive potentials.

In this Letter, we demonstrate trapping of single $^{138}\text{Ba}^{+}$ ions without additional rf confinement in an optical dipole trap at $532\,\text{nm}$ with a lifetime of $166\,\text{ms}$, enabled by efficient optical repumping. This corresponds to an improvement by two orders of magnitude as compared to the state-of-the-art \cite{Huber2014}.
In a near-infrared dipole trap at $1064\,\text{nm}$, the lifetime is further extended to $3\,\text{s}$. Both dipole lasers feature state-dependent optical potentials which we experimentally prepare and characterize.
In addition, we measure the current upper bound for the heating rate in our apparatus caused by the near-infrared laser and ambient fields. 
For the deduced heating rate, sympathetic cooling of the ion and entering the regime of ultracold ion-atom interactions for a variety of mass ratios should be allowed \cite{Tomza2015}.
\begin{figure}[t]
\includegraphics[width = 0.48 \textwidth]{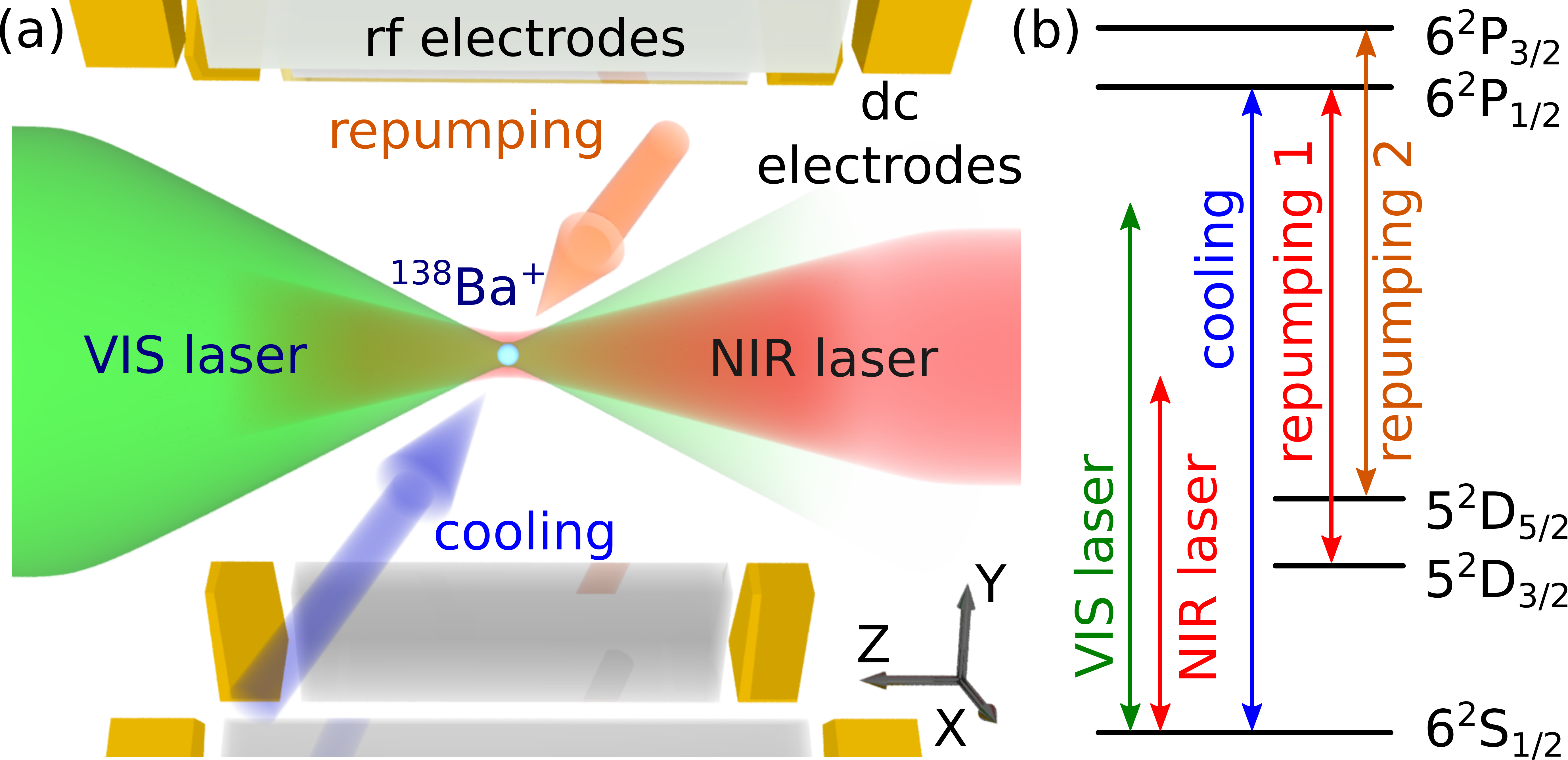}
 \caption{ (color online)
  (a) Setup for optical ion trapping. 
  We prepare a single $^{138}\text{Ba}^{+}$ ion in a segmented linear Paul trap and initialize the ion by Doppler cooling (indicated by the tilted blue arrow) and compensating stray electric fields. 
  We then transfer the ion into either the visible (VIS) or the near-infrared (NIR) dipole trap.
  Both lasers depicted as red (NIR) and green (VIS) Gaussian beams are aligned with the Paul trap's $z$-axis with their foci being centered on the ion.
  During the optical trapping duration $\Delta t_{\text{opt}}$ the rf supply of the Paul trap is disconnected.
  The ion is optionally illuminated with lasers (short orange arrow) used for repumping from the metastable $ D $ level manifolds during $\Delta t_{\text{opt}}$.
  (b) Electronic levels and optical transitions of $^{138}\text{Ba}^{+}$ ions relevant for Doppler cooling at wavelength $\lambda_{\text{cool}} = 493 \, \text{nm} $, repumping ($ \lambda_{\text{rp1}} = 650 \, \text{nm} $ and $ \lambda_{\text{rp2}} = 615 \, \text{nm} $), and optical trapping ($\lambda_{\text{VIS}} = 532\,\text{nm} $ and $\lambda_{\text{NIR}} = 1064\,\text{nm} $).
  The lifetimes of the electron in the $P_{1/2}$ and $P_{3/2}$ levels are $ \tau(P_{1/2}) = 8 \, \text{ns}$ and $ \tau(P_{3/2}) = 6.3 \, \text{ns}$, respectively, with a branching ratio for the decay $P \rightarrow S$ vs $P \rightarrow D$ of approximately $ 3:1 $ \cite{NistData}.\label{fig_scheme}}
\end{figure}

A schematic of our setup is shown in Fig.~\ref{fig_scheme}(a).
We start our experiment by photoionizing \cite{Leschhorn2012} and cooling a single $^{138}\text{Ba}^{+}$ ion to the Doppler limit ($T_{\text{D}} \approx 300\,\mu\text{K}$) in a linear rf trap. 
We prepare the electronic state of the ion in the $S_{1/2}$($D_{3/2}$) manifold by first switching off the cooling (repumping) laser at wavelength $\lambda_{\text{cool}} =  493 \, \text{nm} $ ($\lambda_{\text{rp1}} =  650 \, \text{nm} $), see Fig.~\ref{fig_scheme}(b).
We then transfer the ion adiabatically into either a visible (VIS) or a near-infrared (NIR) dipole trap created by lasers operating at wavelengths $\lambda_{\text{VIS}} = 532\,\text{nm}$ and $\lambda_{\text{NIR}} = 1064\,\text{nm}$, respectively.
This is achieved by turning on the dipole laser while ramping the amplitude of the rf field to zero within $100\,\mu\text{s}$.
Optionally, we turn on additional repumping lasers at wavelengths $ \lambda_{\text{rp1}} = 650 \, \text{nm} $ and $ \lambda_{\text{rp2}} = 615 \, \text{nm} $ tuned to the $D_{3/2} \leftrightarrow P_{1/2}$ and $D_{5/2} \leftrightarrow P_{3/2}$ transitions for fast depopulation of the electronic $D$ manifolds, see Fig.~\ref{fig_scheme}(b).
After a duration $\Delta t_{\text{opt}}$, the rf fields are ramped up again and the cooling and repumping lasers are switched on while the dipole laser is being turned off. 
We then determine with a detection fidelity close to $100\%$ whether the ion has been optically trapped during $\Delta t_{\text{opt}}$ by fluorescence detection on a CCD camera.
The optical trapping probability $p_{\text{opt}}$ is the normalized average of successful optical trapping attempts.

During the initialization and detection stages, radial confinement in the $x$-$y$ plane is generated by rf fields.
Axial confinement along the $z$-axis is currently provided by dc fields, but crossed beams or a standing wave \cite{Enderlein2012} can be used in the future.
The related axial secular frequency ($\omega_z/ 2 \pi \approx 12\,\text{kHz}$) can be considered independent of the nature of the radial confinement and to remain identical for all experiments described in this Letter.
Control over dc fields permits fine-tuning of the axial and radial trapping frequencies and the orientation of the radial principal axes \cite{Schneider2012,Kalis2016}.
This is exploited, e.g., to optimize the cooling process.
Following the preparation of the ion, we compensate stray electric fields by minimizing the radial displacement of the ion caused by switching between two extremal rf confinements ($14\,\text{kHz}\leq\omega_{x,y}^{\text{rf}}/2\pi\leq 140\,\text{kHz}$) \cite{Berkeland1998}.
Using this method, we currently tune $|\vec{E}_{\text{stray}}| \lesssim 10^{-2}\,\text{V/m}$ with a resolution of $10^{-3}\,\text{V/m}$.

Additionally, we spatially overlap one of the two dipole lasers at wavelengths $\lambda_{\text{VIS}}$ and $\lambda_{\text{NIR}}$ with the minimum of the rf field.
Both Gaussian beams are aligned with the axis of the linear Paul trap ($z$-axis).
At the position of the ion, we measure $1/e^2$ intensity waist radii of $w_{\text{VIS}} = (2.6 \pm 0.2)\,\mu\text{m}$ and $w_{\text{NIR}} = (5.2 \pm 0.3)\,\mu\text{m}$, respectively.
The optical powers of the two lasers, $P_{\text{VIS}}\leq 10\,\text{W}$ ($P_{\text{NIR}}\leq 20\,\text{W}$), can be controlled with an acousto-optical modulator and lead to trap depths $U_{\text{VIS}}(S_{1/2})/k_B\leq 120\,\text{mK}$ and $U_{\text{NIR}}(S_{1/2})/k_B\leq 14\,\text{mK}$, where $S_{1/2}$ denotes the electronic state of the ion after preparation and $k_B$ is the Boltzmann constant.
$U_{\text{VIS}}$ ($U_{\text{NIR}}$) is calculated taking into account the influence of residual static electric fields.
The spatial dependence of the intensity of the light field (peak intensities $I_{\text{VIS}} \leq I^{\text{max}}_{\text{VIS}}\approx 1 \times 10^8\,\text{W}\,\text{cm}^{-2}$ and $I_{\text{NIR}} \leq I^{\text{max}}_{\text{NIR}}\approx 5 \times 10^7\,\text{W} \, \text{cm}^{-2}$ respectively) allows to confine the ion in radial directions ($x$ and $y$).
The frequency of the two dipole lasers is far-detuned to the red of the $S_{1/2} \leftrightarrow P_{1/2}$ and $S_{1/2} \leftrightarrow P_{3/2}$ transitions, with detunings in units of linewidth ($\Gamma_{S_{1/2} \leftrightarrow P_{1/2}} = 2\pi\times 15\,\text{MHz}$ and $\Gamma_{S_{1/2} \leftrightarrow P_{3/2}} = 2\pi\times 18\,\text{MHz}$) of $\delta_{\text{VIS}} \approx -2 \times 10^6 \, \Gamma_{S_{1/2} \leftrightarrow P_{1/2}} $ and $\delta_{\text{NIR}} \approx -2 \times 10^7 \, \Gamma_{S_{1/2} \leftrightarrow P_{1/2}} $ \cite{NistData}.
We experimentally determine the radial secular frequencies in the dipole trap by resonant excitation of the ion motion with oscillating electric fields.
At the resonance frequency $\omega_{x,y}^{\text{VIS}}$ ($\omega_{x,y}^{\text{NIR}}$), we detect a decrease in $p_{\text{opt}}$.
Our results agree with the theoretically predicted $U_{\text{VIS}}$ ($U_{\text{NIR}}$) and $\omega_{x,y}^{\text{VIS}}$ ($\omega_{x,y}^{\text{NIR}}$) derived from $I_{\text{VIS}}$ ($I_{\text{NIR}}$), $\delta_{\text{VIS}}$ ($\delta_{\text{NIR}}$) and $w_{\text{VIS}}$ ($w_{\text{NIR}}$) of the dipole lasers \cite{Grimm2000}.
\begin{figure}[t]
  \includegraphics[width = 0.48 \textwidth]{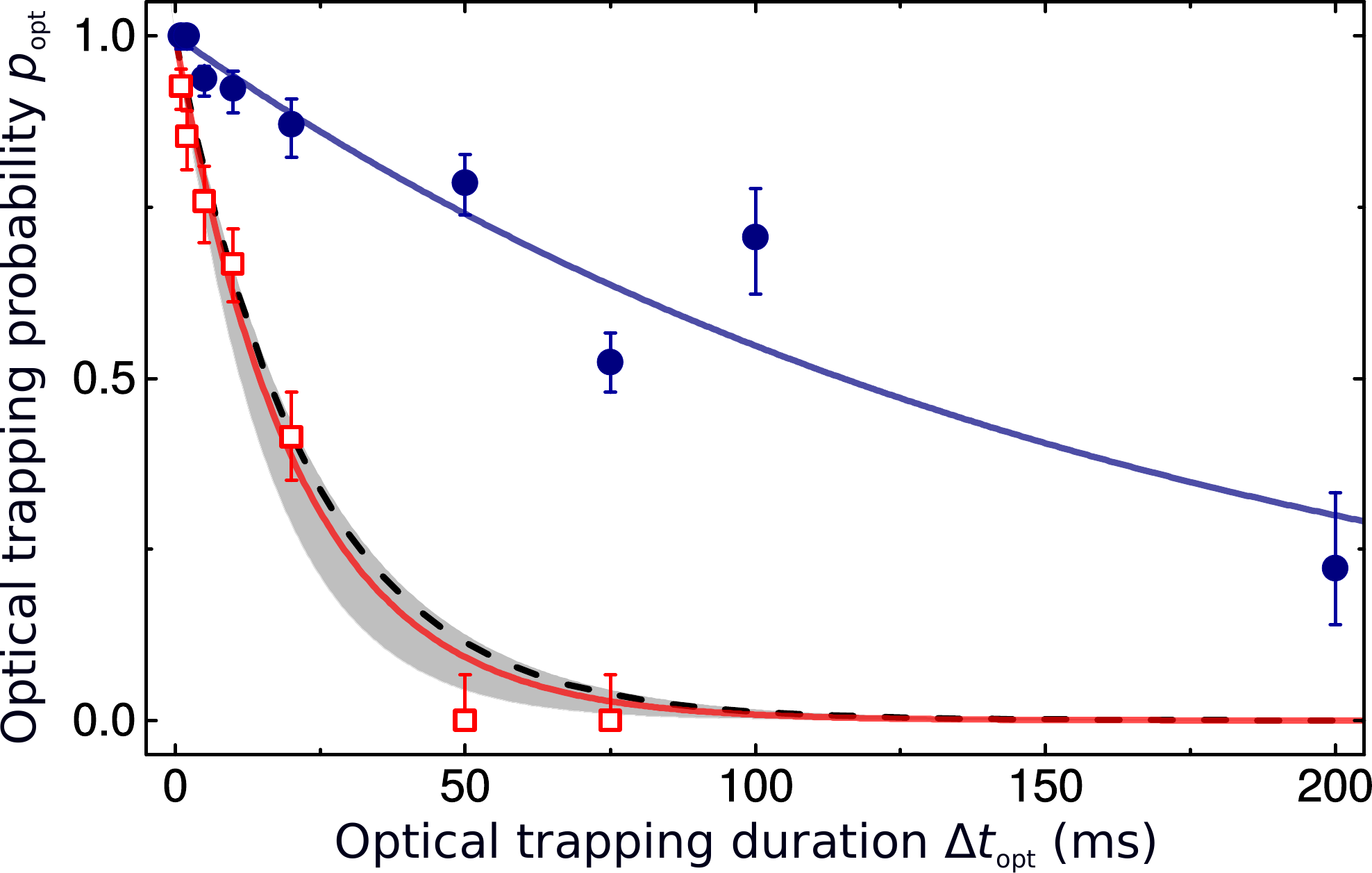}
 \caption{ (color online)
  Lifetime of a single $^{138}\text{Ba}^+$ ion in the VIS dipole trap with and without repumping on the $ D \leftrightarrow P $ transitions.
  We measure the optical trapping probability $p_{\text{opt}}$ in dependence on the trapping duration $\Delta t_{\text{opt}}$.
  Data points represent the mean ratio of successful vs total attempts and the error bars correspond to the $1\sigma$-confidence intervals calculated from the underlying binomial distribution.
  The open squares represent data taken with the dipole laser only [$I_{\text{VIS}}\approx 6 \times 10^6\,\text{W} \, \text{cm}^{-2}$, $U_{\text{VIS}}(S_{1/2})/k_B\approx 8\,\text{mK}$].
  The dashed line shows the decay of $p_{\text{opt}}$ with a lifetime of $\tau_{\text{D}} =(23 \pm 4)\,\text{ms}$ derived from measured off-resonant scattering rates into the $D$ manifolds, where the optical dipole potential is repulsive.
  The shaded area depicts the bounds of a theoretical prediction given our experimental uncertainties, corresponding to a lifetime of $\tau_{\text{theo}} =(20 \pm 4)\,\text{ms}$ (see main text).
  The circles represent data taken with additional repumping lasers (resonant at zero ac Stark shift), focused on the ion during $\Delta t_{\text{opt}}$.
  Solid lines show exponential fits to each dataset.
  The lifetime in the VIS dipole trap, $\tau_{\text{VIS}} = (21 \pm 2)\,\text{ms}$, is increased to $\tau_{\text{VIS,rp}}=(166 \pm 19)\,\text{ms}$ by applying our repumping scheme.\label{fig_lifetime532}
 }
 \end{figure}

We obtain the lifetime of the ion in the VIS dipole trap by measuring  $p_{\text{opt}}$ in dependence on $\Delta t_{\text{opt}}$.
The open squares in Fig.~\ref{fig_lifetime532} represent data taken with the VIS dipole trap at constant $I_{\text{VIS}}= (6.1\pm 0.8) \times 10^6\,\text{W} \, \text{cm}^{-2}$, providing an optical trap depth of $U_{\text{VIS}}(S_{1/2})/k_B\approx 8\,\text{mK}$ (corresponding to $\omega_{x,y}^{\text{VIS}}/2 \pi \approx 80 \,\text{kHz}$).
An exponential fit to the data yields a lifetime of $\tau_{\text{VIS}} = (21 \pm 2)\,\text{ms}$.
For these trapping parameters, we expect an off-resonant scattering rate of $\Gamma_{\text{offr}} = (200 \pm 30)\,\text{Hz}$ \cite{Grimm2000}, corresponding to a photon recoil heating rate of $R_{\text{rec}}\approx 100 \, \mu \text{K/s}$ (with photon recoil energy $E_{\text{rec}} = k_B \times 250\,\text{nK}$).
Independent measurements show that the temperature of the $^{138}\text{Ba}^{+}$ ion, after transfer into the dipole trap, is still on the order of $T_{\text{D}}$.
Thus, the total gain in temperature including recoil heating, $R_{\text{rec}}\cdot \Delta t_{\text{opt}}$ remains an order of magnitude smaller than $U_{\text{VIS}}(S_{1/2})/k_B$ for the investigated trapping durations and does not limit the lifetime.

However, $S_{1/2} \leftrightarrow P_{1/2}$ and $S_{1/2} \leftrightarrow P_{3/2}$ are not closed cycling transitions in $^{138}\text{Ba}^{+}$.
Off-resonant scattering therefore populates metastable $D$ manifolds [see Fig.~\ref{fig_scheme} (b)].
The VIS dipole laser is detuned to the blue with respect to the $D \leftrightarrow P$ transitions and the related optical potentials $U_{\text{VIS}}(D_{3/2})$ and $U_{\text{VIS}}(D_{5/2})$ are repulsive [$(\omega_{x,y}^{\text{VIS}}/2 \pi)^2 \approx - (40\,\text{kHz})^2$].
Within $100\,\mu\text{s}$, the ion is pushed out of the trapping range of any optical potential within $\Delta t_{\text{opt}}$ and rf confinement subsequently applied for detection.
We make use of electron shelving \cite{Nagourney1986} to experimentally investigate scattering into the $D$ manifolds and derive a total rate of $\Gamma_D = (43 \pm 3)\,\text{Hz}$.
The related exponential decay, with a lifetime of $\tau_{\text{D}} =(23 \pm 4)\,\text{ms}$, is shown as the dashed line in Fig.~\ref{fig_lifetime532}.
Our results agree with the theoretically expected lifetime $\tau_{\text{theo}} =(20 \pm 4)\,\text{ms}$, derived from $\Gamma_{\text{offr}}$ and the branching ratios ($ P \rightarrow S$ vs $ P \rightarrow D$) of approximately 3:1 \cite{NistData}, assuming a point like particle centered at the maximal intensity of the dipole laser. 
Taking into account the ion's motion and experimental uncertainties, e.g. in $P_{\text{VIS}}$ and $w_{\text{VIS}}$, yields $1\sigma$-confidence intervals for the predicted $p_{\text{opt}}$ shown as the shaded area.

To counteract this loss mechanism, we apply additional light fields during $\Delta t_{\text{opt}}$ to depopulate the $D$ manifolds.
In order to maintain considerable repumping rates despite the spatially dependent ac Stark shift induced by the dipole lasers, we employ saturation broadening of the transitions by repumping with intensities ($I_{\text{rp1}}$, $I_{\text{rp2}}$) exceeding the saturation intensity $I_{\text{sat}}\approx 10 \text{mW}/\text{cm}^2$ by several orders of magnitude ($I_{\text{rp1,rp2}}/I_{\text{sat}} \approx 10^4$).
With this scheme, we achieve an increase of the lifetime by a factor of 8 to $\tau_{\text{VIS,rp}}=(166 \pm 19)\,\text{ms}$.
However, repumping does not occur instantaneously and scattering events into the $D$ manifolds still increase the mean kinetic energy.
The related heating rate depends on the difference of $U_{\text{VIS}}(S_{1/2})$ and $U_{\text{VIS}}(D_{3/2})$ [$U_{\text{VIS}}(D_{5/2})$] as well as on the randomly distributed dwell time of the ion in the $D$ manifolds.
We expect this effect to dominate the remaining loss rate. Still, one uncertainty stems from current limitations in the alignment procedure of the repumping laser at wavelength $\lambda_{\text{rp2}} =  615 \, \text{nm} $ in our setup.
\begin{figure}[b]
 \includegraphics[width = 0.48 \textwidth]{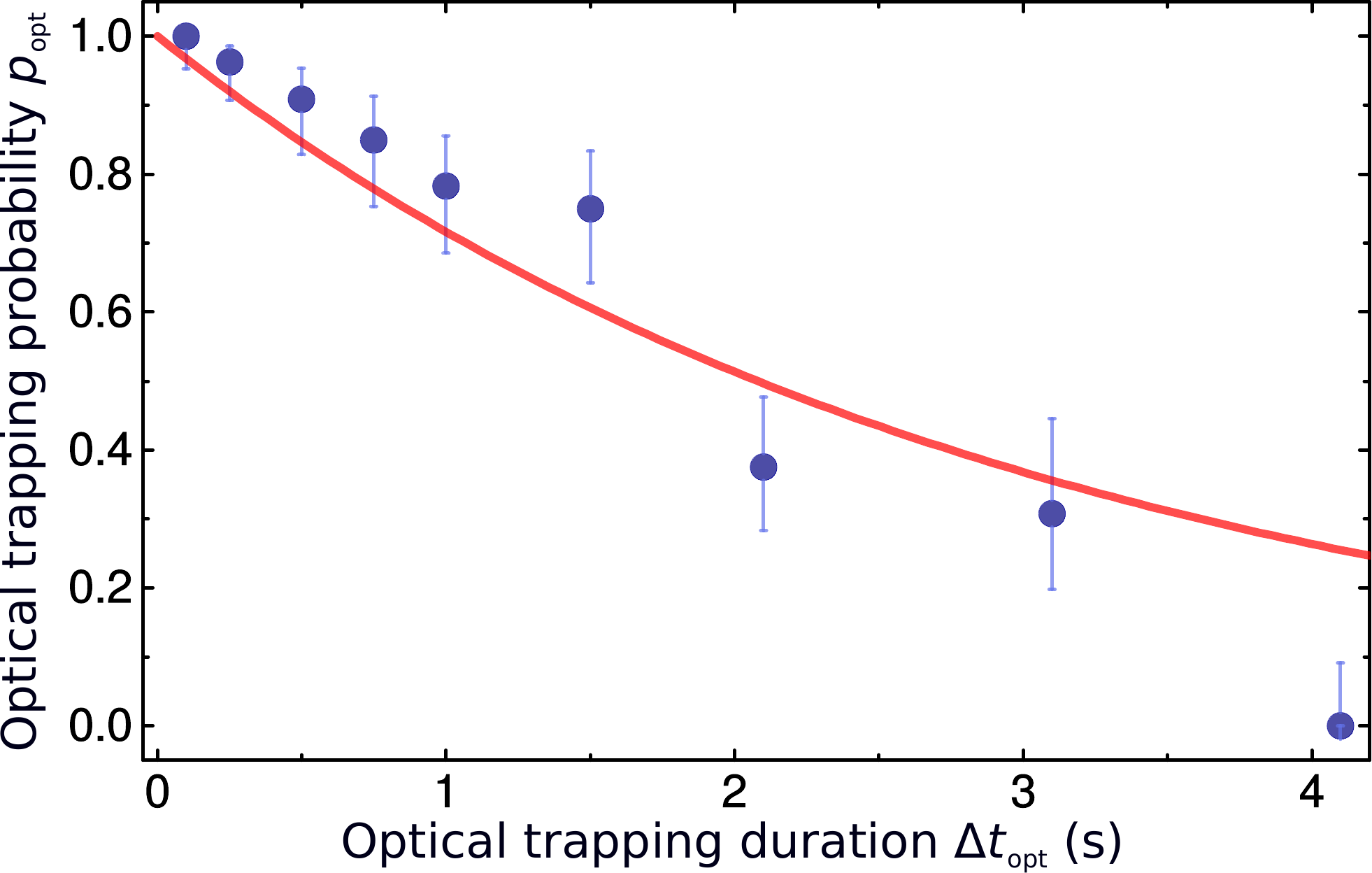}
 \caption{ (color online)
  Optical trapping probability $p_{\text{opt}}$ in dependence on the trapping duration $\Delta t_{\text{opt}}$ for single $^{138}\text{Ba}^+$ ions in the NIR dipole trap. 
  %Data points represent the ratio of successful vs. total attempts and the errorbars correspond to the $1\sigma$-confidence intervals calculated from the underlying binomial distribution \cite{Wilson1927}.
  With a peak intensity $I_{\text{NIR}}\approx 5 \times 10^7\,\text{W}\,\text{cm}^{-2}$, we obtain a trap depth $U_{\text{NIR}}(S_{1/2})/k_B \approx 14\,\text{mK}$.
  The line depicts the result of a fit, assuming exponential decay, yielding a lifetime of $\tau_{\text{NIR}}= (3 \pm 0.3)\,\text{s}$. The error bars of the data points correspond to $1\sigma$-confidence intervals.\label{fig_lifetime1064}
 }
\end{figure}

We further investigate the lifetime of the ion in the NIR dipole trap, which provides confinement in both the $ S $ and $ D $ manifolds.
An intensity of $I_{\text{NIR}} = (4.7 \pm 0.4)\times 10^7\, \text{W}\,\text{cm}^{-2}$ yields an optical trap depth of $U_{\text{NIR}}(S_{1/2})/k_B \approx 14\,\text{mK}$ ($\omega_{x,y}^{\text{NIR}}/ 2 \pi  \approx 60\,\text{kHz}$).
The predicted off-resonant scattering rate $\Gamma_{\text{offr}}=(7.3 \pm 0.6)\,\text{Hz}$ corresponds to a recoil heating rate of $R_{\text{rec}}\approx 3 \, \mu\text{K}/\text{s}$ with a negligible impact on $p_{\text{opt}}$.
With the protocol described above and the ion prepared in the $S_{1/2}$ manifold, we observe that the lifetime of a single $^{138}\text{Ba}^+$ ion in the optical trap reaches $\tau_{\text{NIR}}= (3\pm 0.3)\,\text{s}$, as shown in Fig.~\ref{fig_lifetime1064}.

Similarly to the VIS laser, the optical potential depends on the electronic state, but a photon scattering event into a $D$ manifold only leads to a reduced optical trap depth.
In independent experiments we measure a trap depth of $U_{\text{NIR}}(D_{3/2})/k_B\approx 3\,\text{mK}$, in good agreement with the theoretical prediction \cite{Kaur2015}.
Since $U_{\text{NIR}}(D_{3/2})$ and $U_{\text{NIR}}(D_{5/2})$ amount to approximately $U_{\text{NIR}}(S_{1/2})/5$ and $k_B T_D \approx U_{\text{NIR}}(D_{3/2})/10$, off-resonant scattering into the $D$ manifolds may still dominate the loss rate. Cooling of the ion to sub-Doppler temperatures will allow to substantially reduce the required intensity, further suppressing $\Gamma_{\text{offr}}$.
\begin{figure}[b]
 \includegraphics[width = 0.48 \textwidth]{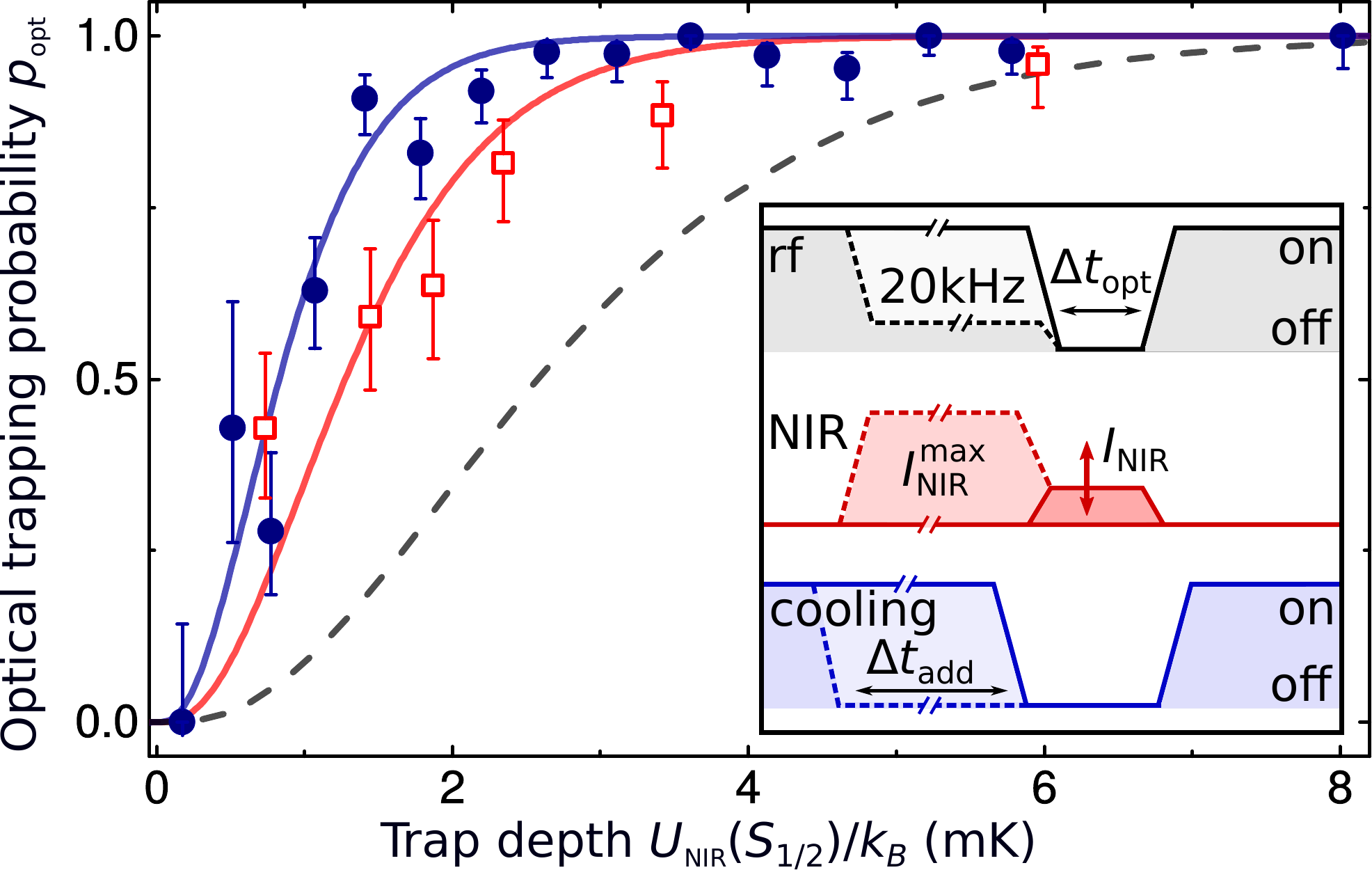}
\caption{ (color online)
  Heating rate for single $^{138}\text{Ba}^+$ ions in an optical dipole trap at ambient fields, analyzing the optical trapping probability $p_{\text{opt}}$ in dependence on trap depth $U_{\text{NIR}}(S_{1/2})$.
  Comparison of $p_{\text{opt}}$ for optical trapping directly after Doppler cooling of the ion (blue circles), and after a delay of $\Delta t_{\text{add}} =500\,\text{ms}$, additionally exposing the ion to ambient and trapping fields (red open squares).
  During $\Delta t_{\text{add}}$, we set the intensity of the dipole laser to $I^{\text{max}}_{\text{NIR}}$, in order to increase the sensitivity to dipole laser fluctuations and apply a weak rf field (see main text).
  We probe the temperature by turning off the residual rf confinement for $\Delta t_{\text{opt}} = 10\,\text{ms}$ and varying $I_{\text{NIR}}$ \cite{Huber2014}. % is chosen to be short compared to the inverse of heating rates. %, but sufficiently long to be sensitive to temperature dependent ion loss out of the optical potential.
  %Data points represent the mean number of successful trapping attempts with $1\sigma$ statistical errors. 
  Solid lines depict fits to the data assuming a radial-cutoff-model \cite{Schneider2012}, yielding temperatures $T_1=(320 \pm 30)\,\mu\text{K}$ and $T_2 = (500 \pm 60)\,\mu\text{K}$. 
  The dashed line shows the theoretical prediction of the radial-cutoff-model for a temperature of $T_3 = 1\,\text{mK}$, emphasizing the sensitivity of the method.
  The error bars of the data points correspond to $1\sigma$-confidence intervals.
  The inset illustrates the two experimental protocols $1$ ($2$) denoted by the solid (dashed) lines corresponding to data points shown as circles (open squares).
  \label{fig_heatrate1064}
  }
\end{figure}

Heating rates of atomic ions could be dominated by effects different from those investigated for neutral atoms in optical traps \cite{Cormick2011}.
For example, ions remain prone to electric field fluctuations.
To estimate an upper bound for the residual heating rate in our apparatus by laser, rf and ambient fields, we extend the protocol by an optional delay, $\Delta t_{\text{add}}= 500\,\text{ms}$, between initialization and optical trapping (inset of Fig.~\ref{fig_heatrate1064}).
During $\Delta t_{\text{add}}$ the dipole laser is set to maximal intensity $I^{\text{max}}_{\text{NIR}}$ while the rf potential is kept negligible compared to the optical confinement ($\omega_{x,y}^{\text{rf}}/2 \pi \approx 20\,\text{kHz}$).
Subsequently we measure $p_{\text{opt}}$ as above without rf confinement in dependence on $I_{\text{NIR}}$ for fixed $\Delta t_{\text{opt}}=10\,\text{ms}$.
Assuming a thermal distribution after initialization as well as after the optional delay, $p_{\text{opt}}$ depends exponentially on the ratio $U_{\text{NIR}}(S_{1/2})/(k_B T_{i})$, where $i=1,2$ denotes the protocols carried out without $(1)$ and with $(2)$ $\Delta t_{\text{add}}$.
Following a radial-cutoff-model \cite{Schneider2012}, truncation of the Boltzmann distribution with varying $U_{\text{NIR}}(S_{1/2}) / k_B \leq 14\,\text{mK}$ reveals an initial temperature of $T_1=(320 \pm 30)\,\mu\text{K}$ and an increase of the temperature to $T_2 = (500 \pm 60)\,\mu\text{K}$ after $\Delta t_{\text{add}}$.
Both datasets and their corresponding fits are shown in Fig.~\ref{fig_heatrate1064}.
The corresponding heating rate amounts to $R_{\text{max}} = (350 \pm 70)\,\mu\text{K}/\text{s}$.
Improvements of the setup, e.g., active stabilization of the intensity and beam pointing and suppression of electric field noise are expected to lower the heating rate substantially.

To summarize, we demonstrate optical trapping of an ion for several seconds in the absence of any radio-frequency field.
A promising approach to reduce the residual loss is to choose an ion species with a more favorable branching ratio from $ P $ states to metastable $ D $ manifolds ($ \text{Yb}^{+}$, $\text{Sr}^{+}$, $\text{Ca}^{+} $ \cite{NistData}) or with a closed-cycling transition ($\text{Be}^{+}$ or $\text{Mg}^{+}$).
An efficient way to reduce the dwell time in the $ D $ states would be to alternate between dipole and resonant repumping lasers at a rate much faster than the ion's motional frequencies $ \omega_{x,y,z}$.
Furthermore, we expect to reduce the required laser intensity for the optical ion trap by up to two orders of magnitude following established cooling methods for state preparation, such as sideband cooling or sympathetic cooling while fully exploiting our current method for the compensation of stray fields.
We note that given the improvement of lifetime demonstrated in this Letter, optical trapping of ions paves the way towards applications currently suffering from fundamental limitations originating from rf micromotion inherent to Paul traps.

One showcase is the study of ion-atom interaction in the quantum regime. 
In these experiments, prolonged interaction times will result in efficient sympathetic cooling rates even at reduced atom densities, which mitigate competing loss mechanisms such as three-body collisions \cite{Kruekow2016}.
This approach relies on the overall ion heating rate to remain sufficiently low compared to the predicted sympathetic cooling rates \cite{Krych2011,Tomza2015}.
In addition, selective control over the potentials of both atoms and ions can be achieved by combining optical traps operated at sufficiently different wavelengths, such as our VIS and NIR dipole lasers. For example, a bichromatic trap will provide deep confinement of alkaline earth ions and a shallow potential for alkali atoms at the same time, since the VIS laser creates a repulsive potential for neutral atom species, such as $\text{Li}$ or $\text{Rb}$.

Long lifetimes are also crucial for scaling to a larger number of ions required for envisioned investigations of structural phase transitions of ion Coulomb crystals confined in a common optical potential.
Higher-dimensional crystals involve ions intrinsically displaced from the rf free axis and are therefore substantially affected by micromotion \cite{Schneider2012b,Thompson2015,Schramm2002}.
In this context, long lifetimes in optical ion traps can enable the combination of state-dependent trapping potentials with the available coherent control of electronic states.
Furthermore, pinning a larger number of ions or atoms in higher-dimensional optical lattices \cite{Zoller2000,Doerk2010} should advance experimental quantum simulations and their scalability, e.g., in the context of frustration governed by quantum spin Hamiltonians, or create novel perspectives such as tunneling or sharing of charges between atoms and ions in a lattice. 

We thank J.~Denter for technical support. This project has received funding from the European Research Council (ERC) under the European Union's Horizon 2020 research and innovation program (grant agreement n$^\circ$ 648330). A.L, J.S., P.W. and M.D. acknowledge support from the DFG within the GRK 2079/1 program, and P.W. additionally thanks the Studienstiftung des deutschen Volkes. L.K. is grateful for financial support from Marie Curie Actions.
\end{document}